\documentclass[twocolumn,linenumbers]{aastex631}

\usepackage{amsmath,amssymb}
\usepackage{graphicx}
\usepackage[utf8]{inputenc}
\usepackage{dcolumn}
\usepackage{bm}
\usepackage{xcolor}
\usepackage{multirow}
\usepackage{array}
\newcolumntype{P}[1]{>{\centering\arraybackslash}p{#1}}

\def \sun	 	{$_{\odot}$}
\def\h{\hat{h}}
\def\d{\delta}

\def\t13{\mathrel{{\theta_{13}}}}
\def\y12{\mathrel{{\tan^2 \theta_{12}}}}
\def\c2{\mathrel{{\chi^2 }}}

\def \sun	 	{$_{\odot}$}

\newcommand{\n}{neutrino}

\newcommand{\be}{\begin{equation}}
\newcommand{\ee}{\end{equation}}
\newcommand{\ba}{\begin{eqnarray}}
\newcommand{\ea}{\end{eqnarray}}

\submitjournal{ApJL}

\shorttitle{}
\shortauthors{Lin et al.}

\begin{document}

\title{Indication of Sharp and Strong Phase-Transitions from NICER Observations}

\author{Zidu Lin}
\email{zlin23@utk.edu}
\affiliation{Department of Physics and Astronomy, University of Tennessee Knoxville,
Knoxville, TN 37996-1200, USA.
}

\author{Andrew. W. Steiner}
\email{awsteiner@utk.edu}
\affiliation{Department of Physics and Astronomy, University of Tennessee Knoxville,
Knoxville, TN 37996-1200, USA.
}
\affiliation{Physics Division, Oak Ridge National Laboratory, TN 37831, USA.}
\begin{abstract}
In this letter, we present a new, weakly model-dependent, test for ``standard" equations of state (EoS) models that disfavor sharp and strong phase-transitions, by using neutron star mass and radius observations. We show the radii of two neutron stars observed by NICER (PSR J0740+6620 and PSR 0030+0451) are correlated if these two neutron stars are built upon standard EoS models. The radii of neutron stars with different masses are sensitive to the pressures at different densities, and the pressures at different densities are strongly correlated in standard EoS models. We further show that the correlation of the neutron star radii can be significantly weakened, when additional degrees of freedom concerning the first-order phase transitions are added into the EoSs. We propose a new quantity, ${D}_{\mathrm{L}}$, which measures the extent to which the linear correlation of the radii of two neutron stars is weakened. Our method gives a 48\% identification probability (with a 5\% false alarm rate) of finding beyond standard EoS models in NICER observations. 
Future observations with higher measurement accuracy can confirm or rule out this identification. Our method is generalizable to any pair of neutron star masses and can be employed with other sets of observations in the future.
\end{abstract}
\section{Introduction}
The composition, as well as the equations of state (EoS) beneath the crust of Neutron Stars (NSs), have long been an open question in both theoretical and experimental nuclear astrophysics. From the theoretical side, the understanding of quantum chromodynamics (QCD) phase diagrams is far from being complete and thus results in large theoretical uncertainties of properties of nuclear matter above the saturation density $n_0$ \citep{Aoki:2006we,Stephanov:2006zvm,Alford:1998sd,Brambilla:2014jmp,Alford:2004pf}. Experimentally, the densities of cold neutron star matter in the cores go well above the saturation density, and producing them at laboratories is inaccessible. Heavy ion collision (HIC) experiments are capable of probing nuclear equation of states at several saturation densities, but the temperatures observed at HIC are much higher than those in neutron star environments \citep{Sorensen:2023zkk}. Astronomical observations provide another way (and probably the most direct way) to study extremely dense matter in neutron stars. In the last six years, three historic observations (GW170817 by LIGO \citep{LIGOScientific:2017vwq}, PSR J0030+0451 by NICER \citep{Riley:2019yda,Miller:2019cac} and PSR J0740+6620 by NICER \citep{Riley:2021pdl,Miller:2021qha}) have significantly strengthened the constraints on neutron star mass-radius (M-R) relationship and thus on EoSs. The impact of NICER and LIGO observations on EoSs has been investigated extensively using both parametric and non-parametric EoS models \citep{Raaijmakers:2021uju,Li:2021thg,Legred:2021hdx,Pang:2021jta,Annala:2021gom,Lattimer:2021emm,Al-Mamun:2020vzu}, using Beyesian inference. However, because of the small number of observed neutron star events and relatively large uncertainties of measured neutron star radii, the M$-$R relationship may have not been constrained at desired accuracy to confirm or rule out the existence of quark matter in the core of neutron stars. Indeed, it is extremely difficult to quantify the possibility of the occurrence of hadron-quark phase transitions in neutron stars using currently available data from astronomical observations. It is not only because of lack of experimental constraints but also because of the poorly understood physics of ultra-dense matter, making it unclear how to construct the prior distribution of EoSs involving hadron quark phase-transition physics. 

The neutron skin thickness of $^{208}\mathrm{Pb}$ and the radii of neutron stars are inherently correlated \citep{Horowitz:2000xj}. Recently, the measurements of parity-violating asymmetry in the elastic scattering of polarized electrons lead to model-independent determination of neutron skin thickness of $^{208}\mathrm{Pb}$ \citep{PREX:2021umo} and $^{48}\mathrm{Ca}$ \citep{CREX:2022kgg}. A strong correlation between the neutron skin thickness of $^{208}\mathrm{Pb}$ and the neutron skin thickness of $^{48}\mathrm{Ca}$ are observed in theoretical nuclear models since both of them are sensitive to the slope of symmetry energy at around saturation densities.

Inspired by this strong neutron skin thickness correlation \citep{Reinhard:2022inh}, it might be interesting to ask: how strong would the correlation of the radii of neutron stars be? Would the radii of PSR J0030+0451 and PSR J0740+6620 measured by NICER agree with the theoretically predicted correlation of NS radii in standard EoS models? We systematically study and further quantify the correlation between the radii of two neutron stars. The existence of strong phase transitions has be related to the difference of radii of two neutron stars \citep{Legred:2021hdx,Essick:2023fso}. We further demonstrate that an \emph{observable} $\mathrm{D}_\mathrm{L}$ obtained from NICER measurements is a better indicator of phase transitions and can be used to test the standard EoS models where the degrees of freedoms concerning the first-order phase transitions are not explicitly included. Indeed, in the absence of other observational evidence, we argue that a strong deviation of two radii from the expected correlation is one of the most sensitive probes of a phase transition.

\section{Formalism}

The radii of neutron stars are mainly determined by the stiffness of EoSs. Given the NS maximum mass, $M_\mathrm{TOV}$ and the low density constraint of EoSs based on chiral effective field theory, one can find the softest(stiffest) possible EoS minimizing(maximizing) the NS radius \citep{Drischler:2020fvz}. As the stiffness of the EoS increases, the size of neutron stars grows \citep{Lattimer:2000nx}. For the correlation between the radii of two neutron stars with mass $M_\mathrm{1}$ and $M_\mathrm{2}$ in a specific type of EoS parametrization, the linear regression of a group of points of theoretically predicted $(R_{\mathrm{M_1}},R_{\mathrm{M_2}})$ with varying stiffness can be approximately characterized by the minimum and maximum radius of these two neutron stars:
\begin{equation} \label{eq:linearRegre}
    R_{\mathrm{M_2}}=R^{\mathrm{min}}_{\mathrm{M_2}}+(R_\mathrm{M_1}-R^{\mathrm{min}}_{\mathrm{M_1}})\frac{R^{\mathrm{max}}_{\mathrm{M_2}}-R^{\mathrm{min}}_{\mathrm{M_2}}}{R^{\mathrm{max}}_{\mathrm{M_1}}-R^{\mathrm{min}}_{\mathrm{M_1}}},
\end{equation}
where $R^{\mathrm{min/max}}_{\mathrm{M_1,M_2}}$ is the minimum and maximum radius of a neutron star with mass ${M}_{1,2}$. Note Eq. \ref{eq:linearRegre} suggests that when $M_1\neq M_2$, the weakening of NS radii correlation (the deviation from the line depicted by Eq. \ref{eq:linearRegre}) may not be best indicated by $R_\mathrm{M_1}-R_\mathrm{M_2}$ used in \citep{Legred:2021hdx,Essick:2023fso}. The lower and upper bound of NS radii have been investigated with minimum model \citep{Drischler:2020fvz}. Using their results, we construct Eq.~\ref{eq:linearRegre}, and the \emph{distance} of a point of $(R_{\mathrm{M_1}},R_{\mathrm{M_2}})$ to this line (which is defined as $D_\mathrm{L} $ in the following ) can be used as a standard to quantify the extent to which the corresponding models (or experimental observations) deviate from the expected correlation, as will be discussed below. In the following, we choose ${M}_1=1.34~\mathrm{M}_\odot$ and ${M}_2=2.0~\mathrm{M}_\odot$, which are close to the mean of the NS mass of PSR J0030+0451 and PSR J0740+6620 measured by NICER. We obtain $R^{\mathrm{min/max}}_{1.34, 2.0}$ from ~\citep{Drischler:2020fvz}, where ${R}^{\mathrm{min}}_{1.34}=9.7\pm\mathrm{0.4}~\mathrm{km}$, ${R}^{\mathrm{min}}_{2.0}=9.0\pm\mathrm{0.0}~\mathrm{km}$, ${R}^{\mathrm{max}}_{1.34}=12.6\pm\mathrm{0.3}~\mathrm{km}$, and ${R}^{\mathrm{max}}_{2.0}=13.1\pm\mathrm{0.2}~\mathrm{km}$, assuming the maximum neutron star mass $M_{\mathrm{TOV}}=2.0~ \mathrm{M}_\odot$. 

\begin{figure*}
     \centering
         \includegraphics[width=0.345\textwidth]{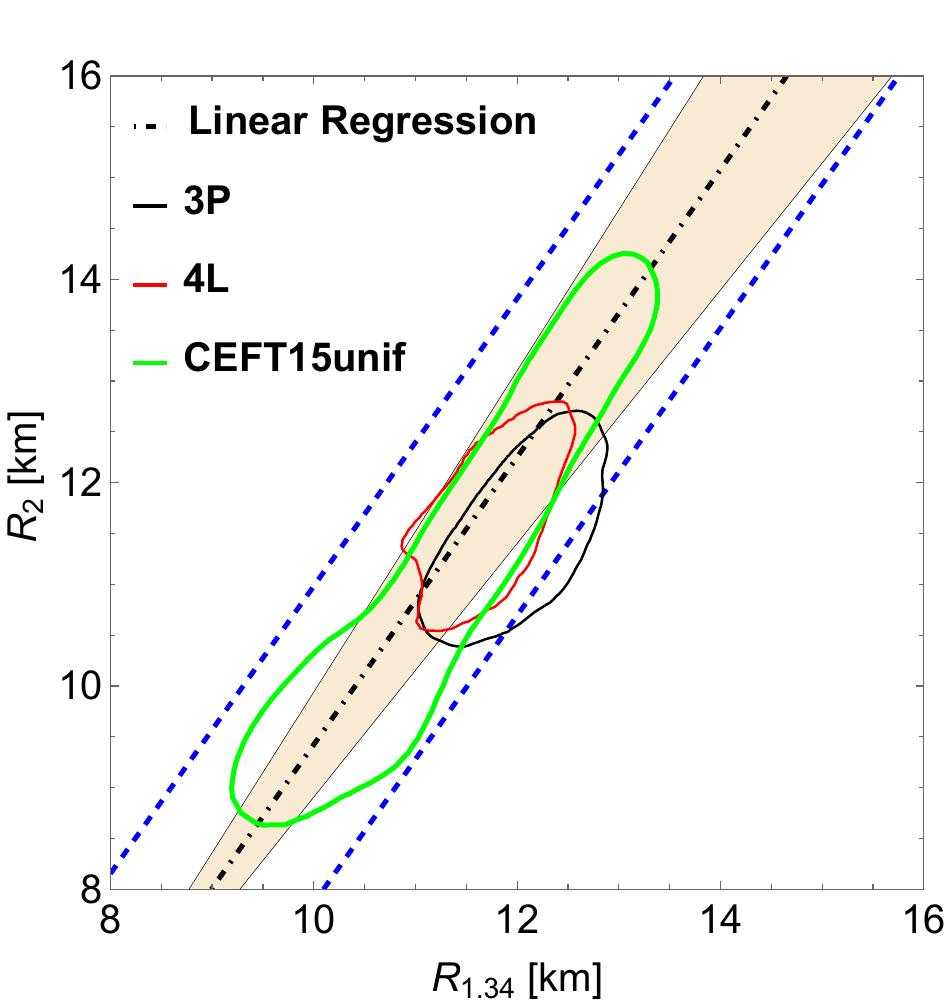}   
         \includegraphics[width=0.45\textwidth]{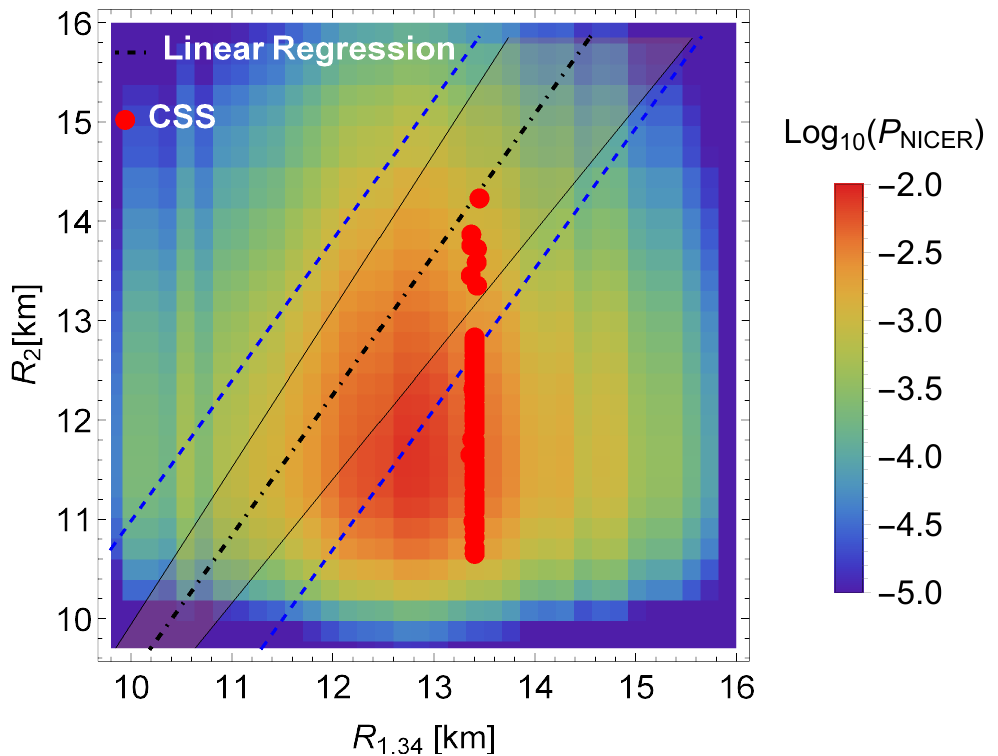} 
        \caption{The line from Eq. \ref{eq:linearRegre} comparing with distributions of $(R_{1.34},R_{2.0} )$ in 3P, 4L and CEFT15unif (left panel) and with the distributions of $(R_{1.34},R_{2.0} )$ in CSS and NICER observations (right panel). The dot dashed black line are described in Eq. \ref{eq:linearRegre} with $M_1=1.34~\mathrm{M}_\odot$ and $M_2=2.0~\mathrm{M}_\odot$ . The black and red solid contour curves in the left panel represent $95\%$ confidential interval of the $(R_{1.34},R_{2.0} )$ based on 3P and 4L EoSs. The green contour, represents the $95\%$ confidential interval of the prior distribution of $(R_{1.34},R_{2.0} )$ based on EoSs published in \citep{Huth:2021bsp}, which are chiral effective field theory EoSs up to 1.5 saturation density and extended with EoSs parametrized using speed of sound. The shaded light yellow band represents the uncertainty of the linear regression line due to the uncertainty of low density chrial effective field theory below two saturation densities (see detailed discussion in \citep{Drischler:2020fvz}). The upper and lower dashed blue lines represent a band of threshold with ${D}_\mathrm{L}^\mathrm{Th}=0.9$ km, beyond which one may claim beyond  standard EoS models are observed (see detailed discussion about the ${D}_\mathrm{L}^\mathrm{Th}$ in text). The colored distribution in the right panel represents the strength of normalized probability distribution of $(R_{1.34},R_{2.0} )$ based on two NICER measurements \citep{Riley:2019yda,Riley:2021pdl}. The red dots represent a selected ensemble of CSS realizations that have similar $\mathrm{R}_{1.34}$ while very different $\mathrm{R}_{2.0}$.  }
        \label{fig:Linear Correlation Compare}
\end{figure*}
\section {Results}
To test the robustness of Eq. \ref{eq:linearRegre}, we present the the prior/posterior probability density distribution of $(R_{\mathrm{M_1}},R_{\mathrm{M_2}})$ of Bayesian inferences based on state-of-the-art standard EoS parametrizations ~\citep{Al-Mamun:2020vzu,Huth:2021bsp}. 
\subsection{The validation of neutron star radii correlation in standard EoS models}
The first EoS model ``3P" from \citep{Al-Mamun:2020vzu}, uses three piecewise continuous polytropes to represent the high-density EoS. Even though the EoS is not differentiable between the three polytropes, this parameterization (together with its prior distribution) mimics the behavior of EoSs without phase transitions or those with only weak phase transitions. The second, ``4L" from \citep{Al-Mamun:2020vzu}, uses four line segments (also piecewise continuous), and this model (together with its prior distribution) mimics the behavior of EoSs without phase transitions, or those with either weak or moderate first order phase transitions. 
The 3P and 4L model, or related parameterizations, have been widely used in Bayesian inference of dense matter EoSs \citep{Annala:2019puf,Steiner:2010fz,Steiner:2011ft,Read:2008iy,Lattimer:2014sga}. The third EoS model ``CEFT15unif" based on chiral effective field theories (CEFT) are introduced in \citep{Huth:2021bsp}. 
The points of $(R_{\mathrm{M1}},R_{\mathrm{M2}})$ from ``3P" and ``4L" models are sampled from the posterior distribution, while the $(R_{\mathrm{M1}},R_{\mathrm{M2}})$ of ``CEFT15unif" are sampled from a much wider prior distribution. 

In the left panel of Fig. \ref{fig:Linear Correlation Compare}, we present the line from Eq.~\ref{eq:linearRegre} as well as the density contour of points of $(R_{1.34},R_{2.0} )$ based on 3P, 4L and CEFT EoS parameterizations. The agreement between the distributions of those points and the model-independent description of neutron star radii correlation in Eq. \ref{eq:linearRegre} is stringkingly good. It suggests that in standard EoS model parameterizations, although the neutron star radii can be easily changed by adjusting the model parameters, their distance to the newly found correlation line in Eq. \ref{eq:linearRegre} cannot be significantly altered. \emph{This makes $D_\mathrm{L}$ an outstanding quantity that characterizes the feature of a large amount of standard EoS model parametrizations}.    
The linear correlations exhibited by points of $(R_{1.34},R_{2.0} )$ of those models can be understood, since in these parameterizations the occurrence of moderate phase transitions are allowed but strong and sharp phase transitions are less likely, and because of that the pressures of NS matter at different density regions are strongly correlated, which finally induces a strong correlation between $R_{1.34}$ and $R_{2.0}$.

\subsection{The deviation from neutron star radii correlation}
We further demonstrate that in model parametrizations where the degrees of freedom regarding the first-order phase transition are introduced, the points of $(R_{1.34},R_{2.0} )$ can significantly deviate from the line of Eq. \ref{eq:linearRegre}, and the $(R_{1.34},R_{2.0} )$ measured by NICER has a non-ignorable possibility to obviously deviate from the correlations, which can not be described by standard EoS parametrizations. 

The Constant Speed of Sound (CSS) parameterization from \citep{Alford:2013aca,Han:2018mtj}, represents either one or two sharp first-order phase transitions. The introduction of sharp phase transitions in CSS model, greatly weakens the correlation between the density-dependent pressures before and after it, and thus likely reduce the correlation between the radii of two neutron stars with unequal mass. This trend has also been observed in recent studies of phase transitions of EoSs using non-parametric EoSs \citep{Essick:2023fso}. 

In the right panel of Fig. \ref{fig:Linear Correlation Compare}, we present the probabilistic distribution of the point of $(R_{1.34},R_{2.0} )$ based on NICER observations. We also show values of $(R_{1.34},R_{2.0})$ from the CSS parameterization, selected to have very similar $R_{1.34}$ but very different $R_{2.0}$. 
The points of $(R_{1.34},R_{2.0} )$ of CSS model can significantly deviate from the line described by Eq. \ref{eq:linearRegre}. This can be understood since a strong and sharp phase transition  
can easily change the pressures that are sensitive to the $R_{2.0}$ while keep the pressures sensitive to $R_{1.34}$ unchanged. Last but not least, the probability density distribution of $(R_{1.34},R_{2.0} )$ based on NICER (denoted as $\mathrm{P}_\mathrm{NICER}$ below) is evaluated by $\mathrm{P}_\mathrm{NICER}(R_{1.34},R_{2.0})=\mathrm{P}_\mathrm{J0030+0451}(R_{1.34})\times \mathrm{P}_\mathrm{J0740+6620}(R_{2.0})$, where $\mathrm{P}_\mathrm{J0030+0451}(R_{1.34})$ and $\mathrm{P}_\mathrm{J0740+6620}(R_{2.0})$ stand for the probability that a neutron star with 1.34 $\mathrm{M}_\odot$ (2.0 $\mathrm{M}_\odot$) have radius of $\mathrm{R}_{1.34}$ ($\mathrm{R}_{2.0}$) based on observation of PSR J0030+0451 (PSR J0740+6620). 

The important observation is that the probability
distribution of the $(R_{1.34},R_{2.0} )$ point inferred by NICER
has a significant statistical weight far from the expected linear correlation. This region cannot be easily reached by the 3P and 4L models, but can be easily reached by CSS models.

\subsection{The weakly-model-dependent test for standard EoS models}\label{sec:DL test}

\begin{figure}
     \centering         \includegraphics[width=0.45\textwidth]{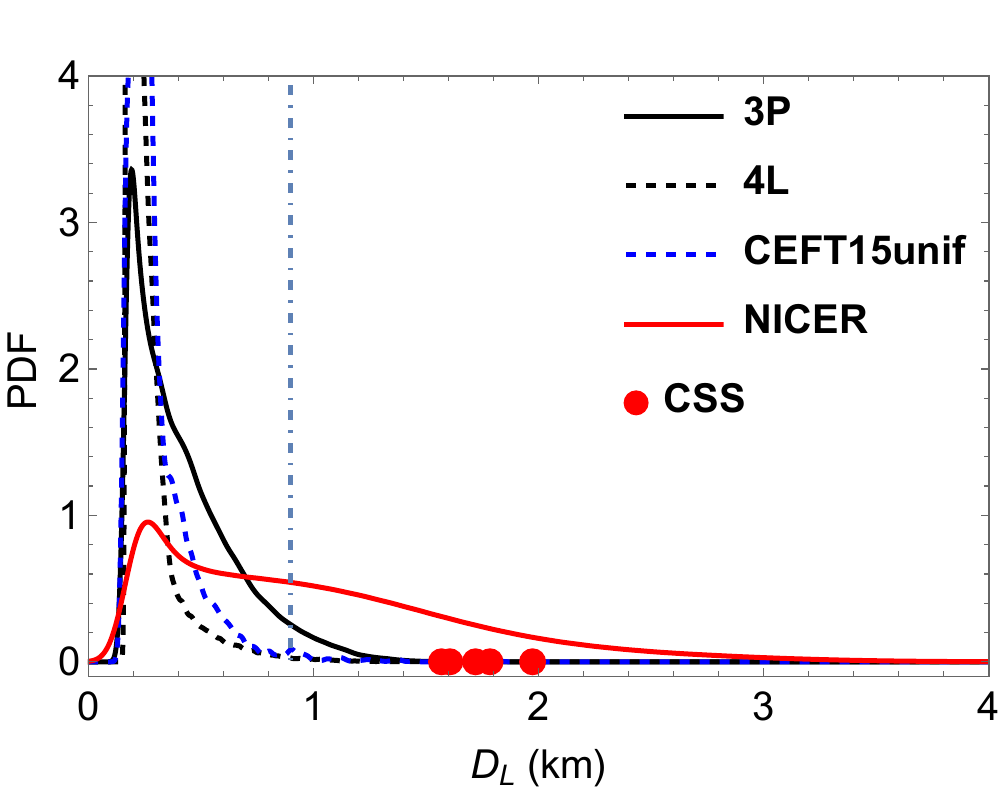}   
        \caption{The distribution of $\{{D}_\mathrm{L} \}$ in 3P (black solid), 4L (black dashed), CEFT15unif (blue dashed) and in NICER observations. The five red dots represent a selected ensemble of CSS realizations that have ${D}_\mathrm{L}>1.5$ km that lie well beyond the distribution of $\{{D}_\mathrm{L} \}$ in 3P and 4L. The blue vertical dotdashed line represent ${D}_\mathrm{L}^\mathrm{Th}=0.9$ km. The data of selected EoS models are described in Zenodo (https://zenodo.org/doi/10.5281/zenodo.13700330)}
        \label{fig:Linear Correlation Distance}
\end{figure}

 Since the line characterizing the correlation of radii of two neutron stars in Eq. \ref{eq:linearRegre} is weakly model-dependent, we use $D_L$ to measure the extent to which the radii do not match the expected correlation. In Fig. \ref{fig:Linear Correlation Distance}, based on the prior/posterior distribution of points of $(R_{1.34},R_{2.0} )$, we have $\mathcal{P} ({D}^\mathrm{3P}_\mathrm{L})$,  $\mathcal{P} ({D}^\mathrm{4L}_\mathrm{L})$ and $\mathcal{P} ({D}^\mathrm{CEFT}_\mathrm{L})$ which give the probability distributions of ${D}_\mathrm{L}$ based on 3P, 4L and Chiral EFT models from \citep{Huth:2021bsp}. The $\mathcal{P} ({D}^\mathrm{NICER}_\mathrm{L})$, obtained from probabilistic interpretation of two NICER measurements, are generated and compared with the former. The distribution of $\mathcal{P} ({D}^\mathrm{NICER}_\mathrm{L})$ has an obviously larger spread comparing to  $\mathcal{P} ({D}^\mathrm{3P}_\mathrm{L})$, $\mathcal{P} ({D}^\mathrm{4L}_\mathrm{L})$ and $\mathcal{P} ({D}^\mathrm{CEFT}_\mathrm{L})$.

To quantitatively estimate the probability that scenarios beyond standard EoS parameterizations occur in NICER observations, we calculate:
\begin{equation}\label{eq:PI}
    \mathrm{P}_{\mathrm{I}}=\int_{{D}^{\mathrm{Th}}_{\mathrm{L}}}^\infty \mathcal{P} ({D}^\mathrm{NICER}_\mathrm{L}) d {D}_\mathrm{L}~,
\end{equation}
and 
\begin{equation}\label{eq:PFA}
    \mathrm{P}_{\mathrm{FA}}=\int_{{D}^{\mathrm{Th}}_{\mathrm{L}}}^\infty \mathcal{P} ({D}^\mathrm{standard}_\mathrm{L}) d {D}_\mathrm{L}~,
\end{equation}
where $\mathrm{P}_{\mathrm{I}}$ stands for the identification probability of EoSs beyond standard scenario and $\mathrm{P}_{\mathrm{FA}}$  is the false alarm rate. The concept of $\mathrm{P}_\mathrm{I}$ and $\mathrm{P}_\mathrm{FA}$ has been widely used in statistical signal processing and detection theory \citep{2009fundamentals}. We define ${D}^{\mathrm{Th}}_\mathrm{L}$ to be a threshold of ${D}_\mathrm{L}$, above which one can claim that an EoS beyond a ``standard scenario" exists.
The $\mathcal{P} ({D}^\mathrm{NICER}_\mathrm{L})$ is the the probability distribution of ${D}_\mathrm{L}$ based on NICER observations and the $\mathcal{P} ({D}^\mathrm{Standard}_\mathrm{L})$ is the probability distribution based on ``standard'' EoS models (such as 3P, 4L, CEFT15unif and their associated probabilities obtained from Bayesian analysis). Note the choice of ${D}^{\mathrm{Th}}_\mathrm{L}$ depends on $\mathcal{P} ({D}^\mathrm{Standard}_\mathrm{L})$ as well as the desired false alarm rate (the value of $\mathrm{P}_\mathrm{FA}$) of a measurement. The standard models analyzed in this paper include a large variety of parameterized models that have no phase-transitions or do not have phase transitions with large regions of density having a very small or zero sound speed. We do not expect including more standard model parameterizations will qualitatively change the range of $\mathcal{P} ({D}^\mathrm{standard}_\mathrm{L})$. Indeed, the distribution of ${D}_\mathrm{L}$ of both 3P, 4L and CEFT models vanish when ${D}_\mathrm{L}>1.5$ km. 

In Fig. \ref{fig:Linear Correlation Distance}, we then get $\mathrm{P}_{\mathrm{I}}=48\%$ using Eq. \ref{eq:PI} and \ref{eq:PFA}, when the ${D}_\mathrm{L}^\mathrm{Th}=0.9$ km is chosen. The choice of ${D}_\mathrm{L}^\mathrm{Th}=0.9$ (represented by a vertical dot-dashed line of this figure) corresponds to $\mathrm{P}_{\mathrm{FA}}=4.7\%$~($0.5\%$ or $1.2\%$) when the ``standard scenario" is represented by 3P (4L or CEFT) model. There could be realizations of NS EoSs with first-order-phase transitions that have $D_\mathrm{L}<0.9~\mathrm{km}$ and realizations of non-CSS models that have $D_\mathrm{L}>0.9~\mathrm{km}$ (due to unclear physics of high-density EoS models). Consequently, we choose to conservatively interpret $\mathrm{P}_\mathrm{I}$ of NICER observations as the occurrence probability of beyond standard EoS parameterizations, rather than an indication of CSS model. Also, EoS models with phase transitions cannot be ruled out by observation of $D_\mathrm{L}<{D}_\mathrm{L}^\mathrm{Th}$. 

\subsection{The future application of $\mathrm{D}_\mathrm{L}$ test on the presence of first-order phase transitions}

Partly due to the different emitting spot models, the neutron star mass and radii of PSR J0030+0451 and PSR J0740+6620 reported in \citep{Riley:2019yda, Riley:2021pdl, Miller:2019cac, Miller:2021qha} have different central values and uncertainties. And the reported radii from NICER measurements all have uncertainties $\gtrsim 1.0~ \mathrm{km}$. In the $D_\mathrm{L}$ test introduced in Sec. \ref{sec:DL test}, the probability of identifying EoS beyond standard scenarios sensitively depends on the uncertainties of measured neutron star radii. And it is important to understand the performance of this test with a better measurement accuracy of NS radii. Recently, a new NICER analysis combined with chiral effective field theories \citep{2024arXiv240706790R} demonstrated the possibility of greatly reducing the uncertainty of NS radii by applying joint constraints. Considering the uncertainty of the neutron star mass and radii in the current state and the possible future NICER measurements with improved accuracy, we discuss three sets of mock data of neutron star radii and their corresponding identification probability of EoSs beyond standard scenario based on the $\mathrm{D}_\mathrm{L}$ test. The mock data are generated based on the measurements from NICER and are summarized in Tab. \ref{tab: mock table}. In mock data A, the mean of $\mathrm{R}_{1.34}$ and $\mathrm{R}_{2.0}$ are chosen to be close to the central value of current NICER measurements \citep{Riley:2019yda, Riley:2021pdl}. In mock data B, the mean of $\mathrm{R}_{1.34}$ is 1.1 km larger (which is approximately the upper bound of the neutron star radius of PSR J0030+0451) than the one in mock data A and $\mathrm{R}_{2.0}$ is 1.0 km smaller (which is approximately the lower bound of the neutron star radius of PSR J0740+6620) than the one in mock data A. In mock data C, the mean of $\mathrm{R}_{1.34}$ is 1.1 km smaller (which is approximately the lower bound of the neutron star radius PSR J0030+0451) than the one in mock data A and $\mathrm{R}_{2.0}$ is 1.0 km larger (which is approximately the upper bound of the neutron star radius PSR J0740+6620) than the one in mock data A. In all three sets of the mock data, the uncertainty of the neutron star radii is assumed to be 0.5 km, which is about half of the uncertainties of the current NICER measurements and is slighly smaller than the uncertainties reported in \citep{2024arXiv240706790R}.

An interesting finding is that based on mock data B, the standard EoS scenario is very likely to be ruled out, while in the scenario similar as mock data A or C, the likelihood of ruling out standard EoS parameterizations is smaller than the $\mathrm{P}_\mathrm{I}=48\%$ based on the NICER data in \citep{Riley:2019yda, Riley:2021pdl}. Pairs of neutron star radii $(\mathrm{R}_{1.34}, \mathrm{R}_{2.0})$ that are closer to the threshold (defined by $\mathrm{D}_\mathrm{L}^\mathrm{Th}$ and illustrated by the upper and lower blue dashed lines in Fig. \ref{fig:Linear Correlation Compare}) imply that a higher measurement accuracy is needed to identify the beyond-standard scenarios. Conversely, for pairs further away from the threshold, a relatively low measurement accuracy would be enough to identify the beyond-standard scenarios. This method provides a general description of desired accuracy for identifying beyond-standard EoSs, which depends on the central value of measured NS radii.

\begin{table}[h]
\caption{\label{tab:X1} Mock NICER data of the radius of 1.34 and 2.0 solar mass neutron stars and also the $P_\mathrm{I}$ when $D_\mathrm{L}^\mathrm{Th}=0.9~\mathrm{km}$. See text for detailed explanation of the choice of mock data. }\label{tab: mock table}
\centering{}%
\setlength{\tabcolsep}{0.2cm}
\begin{tabular}{cccc}
\hline
\hline
  Mock Data  & $\mathrm{R}_{1.34}$ (km)  & $\mathrm{R}_{2.0}$ (km) & $P_\mathrm{I}$ \tabularnewline
\hline
 A & $12.8\pm 0.5$ & $12.2\pm 0.5$ & 36.6\% \tabularnewline
 B & $13.9\pm 0.5$ & $11.2\pm 0.5$ & 99\% \tabularnewline
 C & $11.7\pm 0.5$ & $13.2\pm 0.5$ & 41.1\% \tabularnewline
 
\hline
\hline
\end{tabular}
\end{table}

\section{Conclusions and Discussion}
The presence of a phase transition is an important diagnostic of QCD. Usually, the identification of phase transitions in neutron stars is performed via reconstructions of full NS M-R curves and EoSs. However, phase transitions are local feature of EoSs. The identification of high-density EoS features relys on the detailed structure of a NS M-R curve, which is difficult to be reconstructed given our current observations that have only a few measurements of NS radii with relatively large uncertainties. 

This letter introduces a new method to identify phase transitions by investigating the two-point correlation of neutron star radii. We found the point $(R_{1.34},R_{2.0} )$ based on NICER data may significantly deviate from a correlation line of NS radii. If this is the case, it can be difficult to explain the NICER data using standard EoS models with weak or non-existent phase transitions. We quantitatively estimate the identification probability of EoSs beyond standard scenario with false identification probability $\mathrm{P}_\mathrm{FA}\leq5\%$. The full identification performance with varying $D_\mathrm{L}^\mathrm{Th}$ can be quantified by the receiver operating characteristic (ROC) curve and the area under this curve (AUC). The model-dependence of this method comes from an imperfect understanding of low-density EoSs that influences the constraints for maximum and minimum radii of NSs. It also results from the sampled points of $(R_{1.34},R_{2.0} )$ coming from theoretical EoS mdoels as representative of the ``standard scenario''. Given a large ensemble of EoSs (534,856 4L EoSs, 436,875 3P EoSs, and EoSs in \citep{Huth:2021bsp}) we have tested, this model dependence is likely weak. Comparing to traditional method of identifying the probability of phase-transition-occurrence by calculating the Bayesian factor, this new method \emph{does not} need to construct EoS parameterizations with phase transitions, whose form and degrees of freedom are highly uncertain. 

As the gap between two NS masses decreases, the correlation between the NS radii becomes stronger, and an identification of phase transitions using this method becomes more accurate. Indeed, when two neutron stars have equal mass, even a slight deviation from the linear correlation clearly indicates a first-order phase transition. Since it means two neutron stars with equal mass but unequal radii (the ``twin-star'' scenario proposed by \citep{Benic:2014jia,Mishustin:2002xe,Alford:2017qgh}) are observed. The correlation of NS radii is actually a generalized ``twin star scenario" where the two neutron stars have unequal masses, which have not been systematically studied and this work makes the first step and a roadmap for the next ones. Also, with number of N neutron stars observed, N(N-1)/2 number of two-point correlations can be investigated. The $D_\mathrm{L}$ test is only sensitive to the phase-transitions at a specific range of densities (approximately from the central density of the lighter NS to the central density of the heavier NS analyzed by this method). The Nth newly observed NS brings new chances of identifying the phase transitions that may occur at densities not sensitive to current NICER measurements. In the appendix, we discuss more quantitative details to illustrate how the correlation strength varies with mass differences and how the future phase transition detection would be be influenced by it. We also illustrate the dependence of phase transition detection on the phase transition densities. In the future, the observations from next-generation gravitational wave detectors \citep{Maggiore:2019uih,Evans:2021gyd, LIGOScientific:2014pky,VIRGO:2014yos} may provide a next-to-leading-order measurement of the tidal deformability $\delta \tilde{\Lambda}$ from binary neutron star mergers \citep{Chatziioannou:2021tdi}, which enables one to study the mass and the radii of two neutron stars involved in a merger system separately. Those future observations could be pivotal for phase-transition identifications.

To conclude, we find the combined NICER observations (PSR J0740+6620 and PSR 0030+045) can deviate from the theoretically predicted correalations of NS radii based on standard EoS models with non-vanishing statistical significance. Our linear correlation method (1) quantitatively estimates the occurrence probability of models beyond the standard scenario with weak model dependence and (2) can be significantly more effective when additional NS radii are measured by next-generation gravitational wave and X-ray detectors. Finally, we note that recently there is an updated mass-radius analysis of the 2017-2018 NICER data set of PSR J0030+0451 \citep{Vinciguerra:2023qxq}, and we plan to include updated NICER data using this method in future investigations.           

\begin{acknowledgements}

We thank A. Schwenk and I. Tews for kindly providing the data of EoSs in \citep{Huth:2021bsp} and fruitful discussions with Sophia Han and Shuzhe Shi. 
ZL and AWS were supported by NSF PHY 21-16686. AWS was also supported by the Department of Energy Office of Nuclear Physics.

\end{acknowledgements}

\software{
\emph{O2scl}, Wolfram Mathematica
}

\appendix
\section{Neutron star correlations with various neutron star mass gap}

\begin{figure*}
     \centering
         \includegraphics[width=0.45\textwidth]{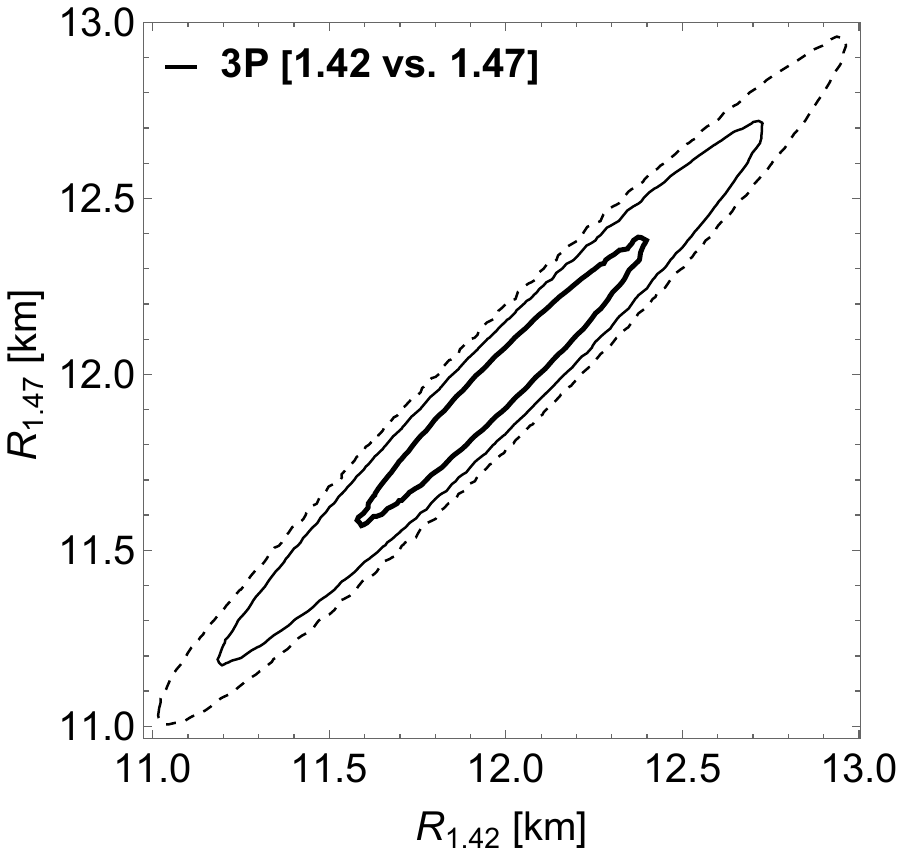}   
         \includegraphics[width=0.43\textwidth]{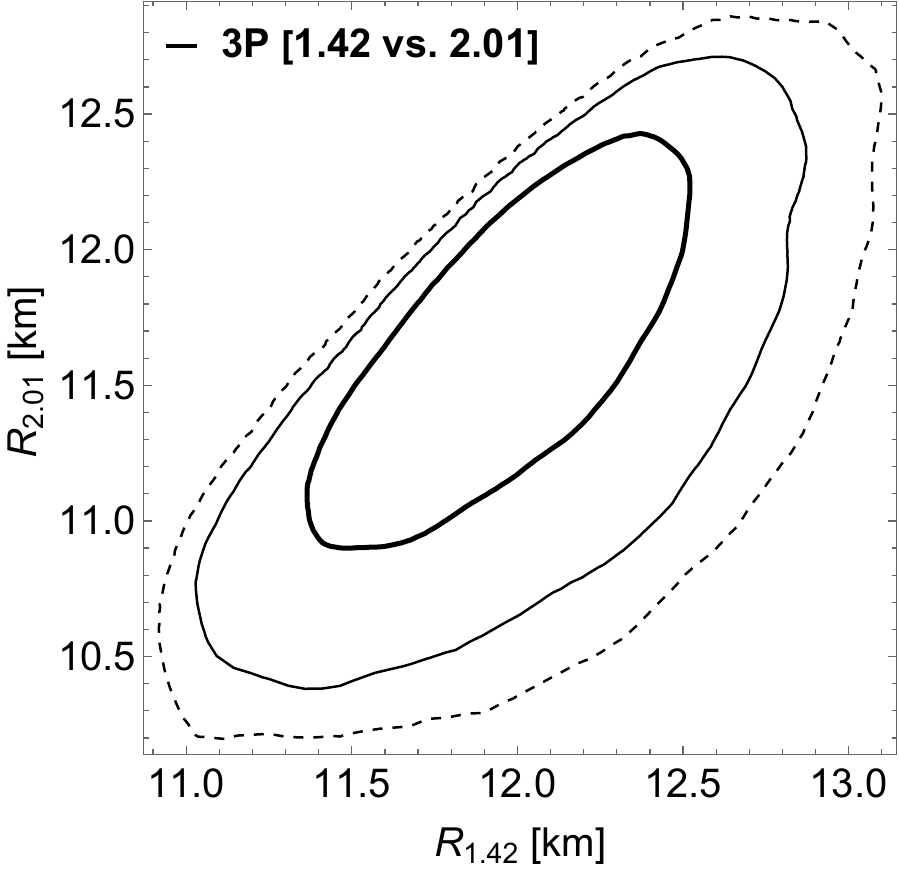} 
         \includegraphics[width=0.43\textwidth]{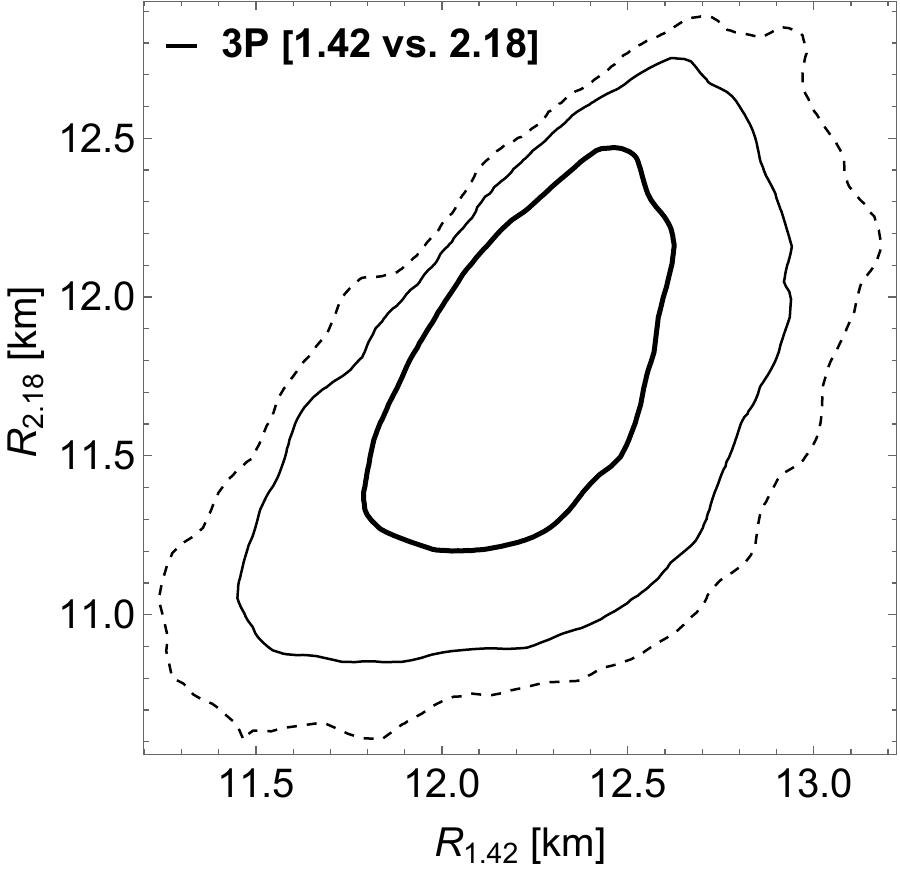} 
        \caption{We present the correlation between two neutron star radii in three different pairs-- $R_{1.47} ~\mathrm{Vs} ~R_{1.42}$ (upper left), $R_{2.01} ~\mathrm{Vs} ~R_{1.42}$ (upper right) and $R_{2.18} ~\mathrm{Vs} ~R_{1.42}$ (lower middle) based on the 3P model. The Pearson correlation between the neutron star radii are 0.99 (upper left), 0.67 (upper right) and 0.54 (bottom) respectively. The solid thick, solid thin and dashed thin curves correspond to the one sigma, two sigma and three sigma contours of the density distribution.}
        \label{fig:Linear Correlation different pair}
\end{figure*}

In Fig. \ref{fig:Linear Correlation different pair} , the correlation between neutron star radii in three different neutron star pairs are shown. As expected, the correlation between the radii of two neutron stars becomes significantly stronger as the mass gap decreases. The phase transition detection accuracy depends not only on the expected neutron star radii correlation strength shown in Fig. \ref{fig:Linear Correlation different pair}, but also on the measurement uncertainty of future NICER observations. It would be extremely interesting to perform quantified analysis using this method when new NICER measurements with different neutron star mass is available.   

As the mass gap approaches to zero ($\mathrm{M}_1\approx \mathrm{M}_2$), the slope of eq. \ref{eq:linearRegre} in the manuscript approaches to 1 and the intercept of eq. \ref{eq:linearRegre} approaches to 0. Consequently, it simply reduces to $R_\mathrm{M_1}=R_\mathrm{M_2}$. In this case, the $D_\mathrm{L}$ introduced in this method reduces to $\Delta R=R_\mathrm{M_1}-R_\mathrm{M_2}$, which is a commonly discussed quantify in previous works concerning the phase-transition possibilities. In this method, the mass gap effect was systematically considered when using the NS radii information to study the phase-transition possibilities.  

\begin{figure*}
     \centering
         \includegraphics[width=0.45\textwidth]{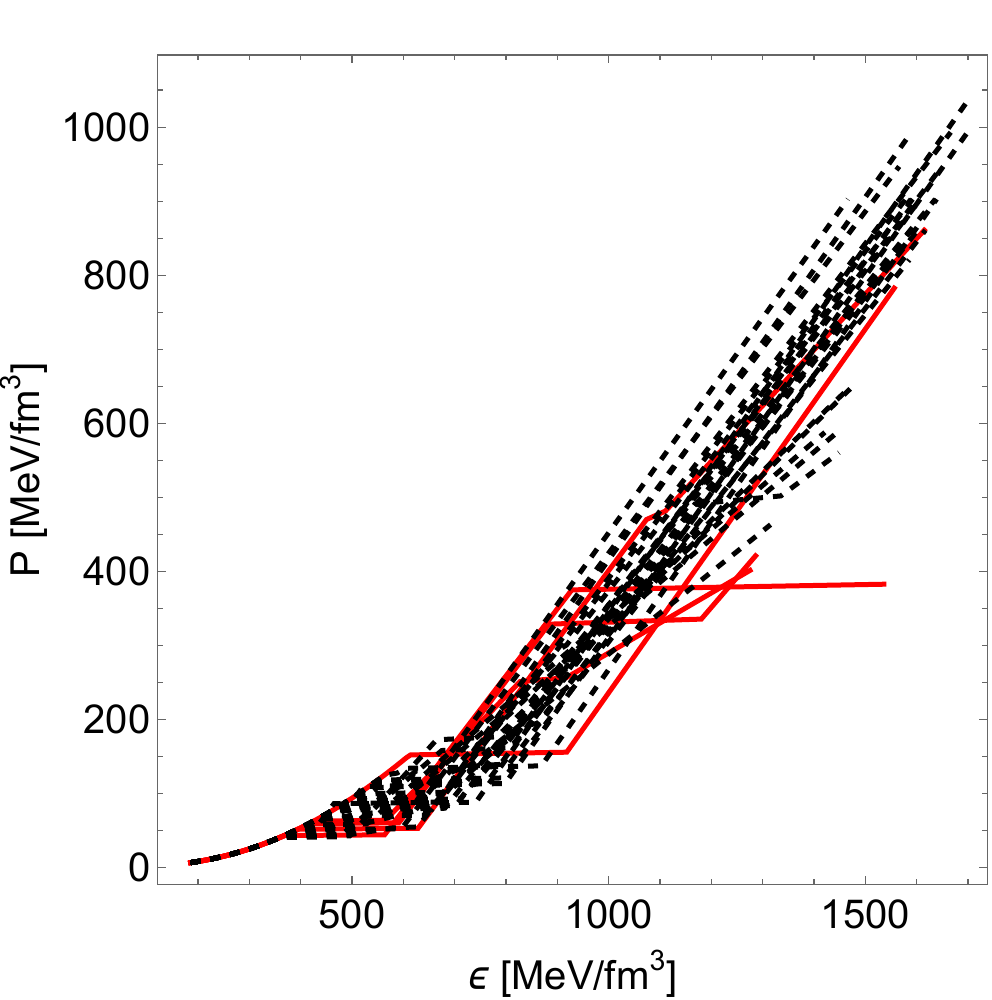}   
         \includegraphics[width=0.43\textwidth]{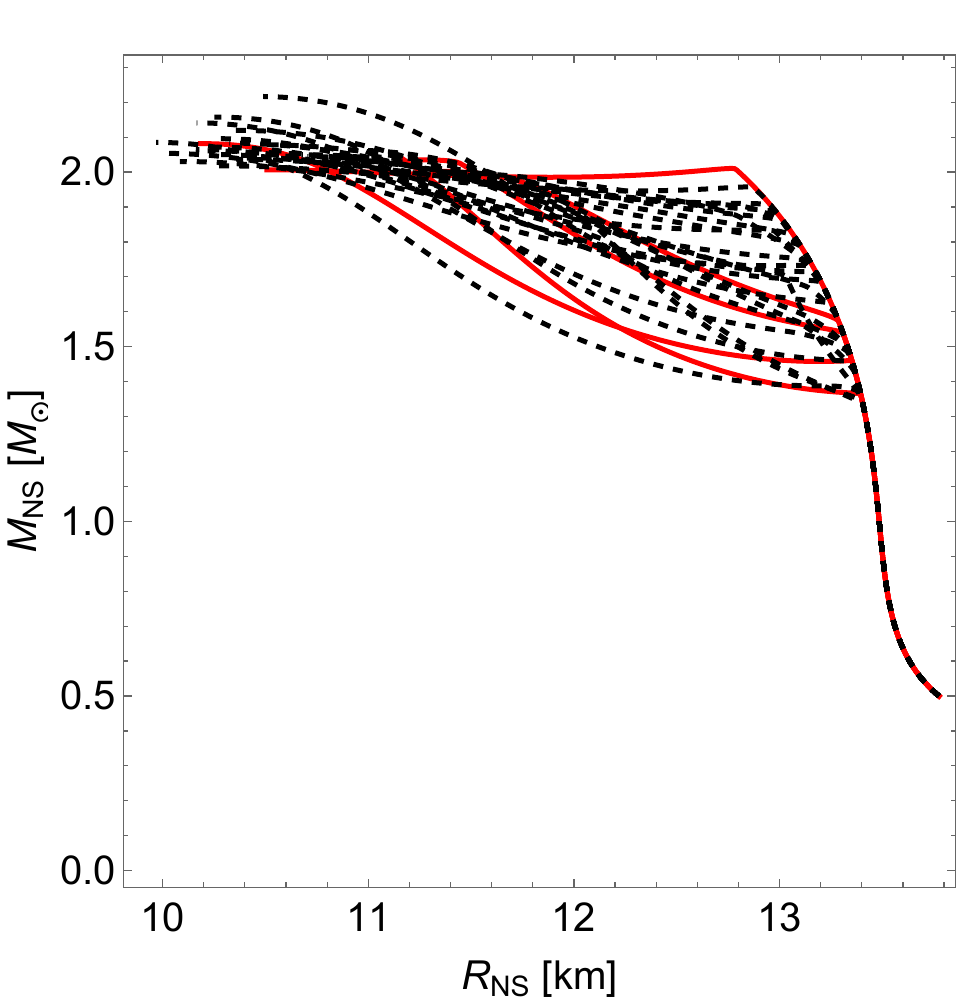} \includegraphics[width=0.43\textwidth]{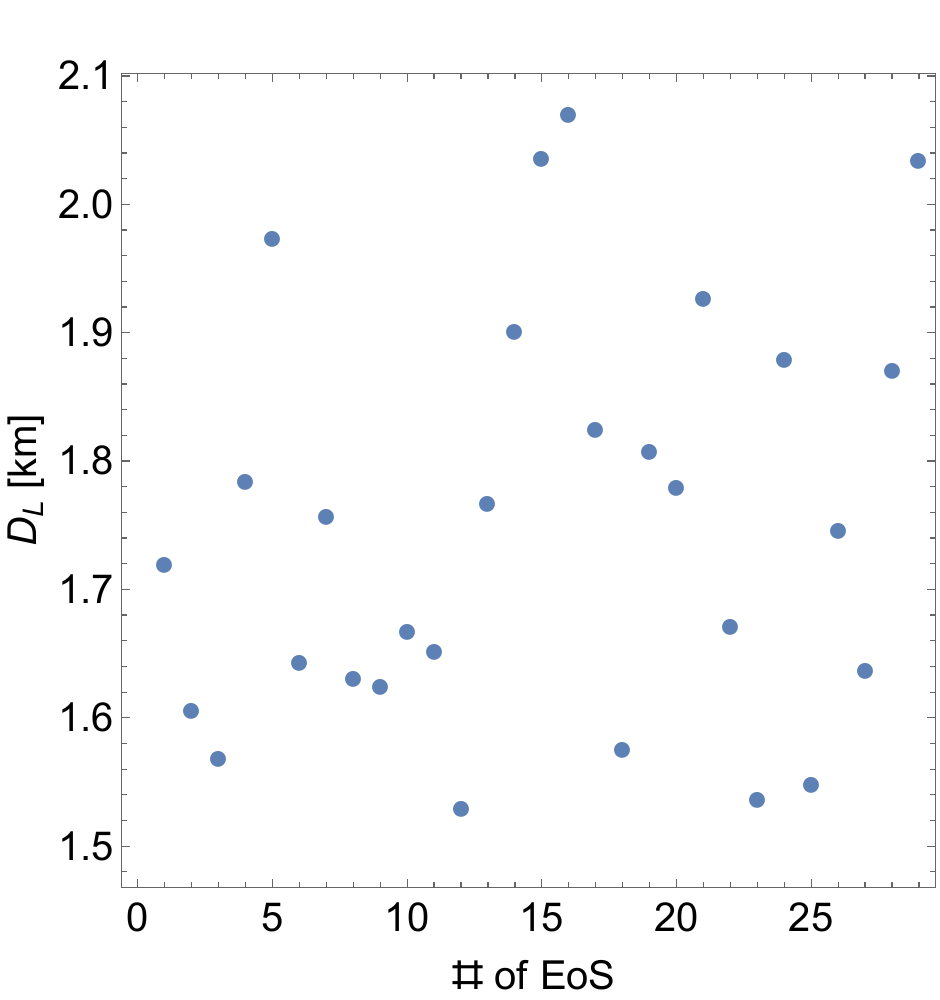}
         \includegraphics[width=0.43\textwidth]{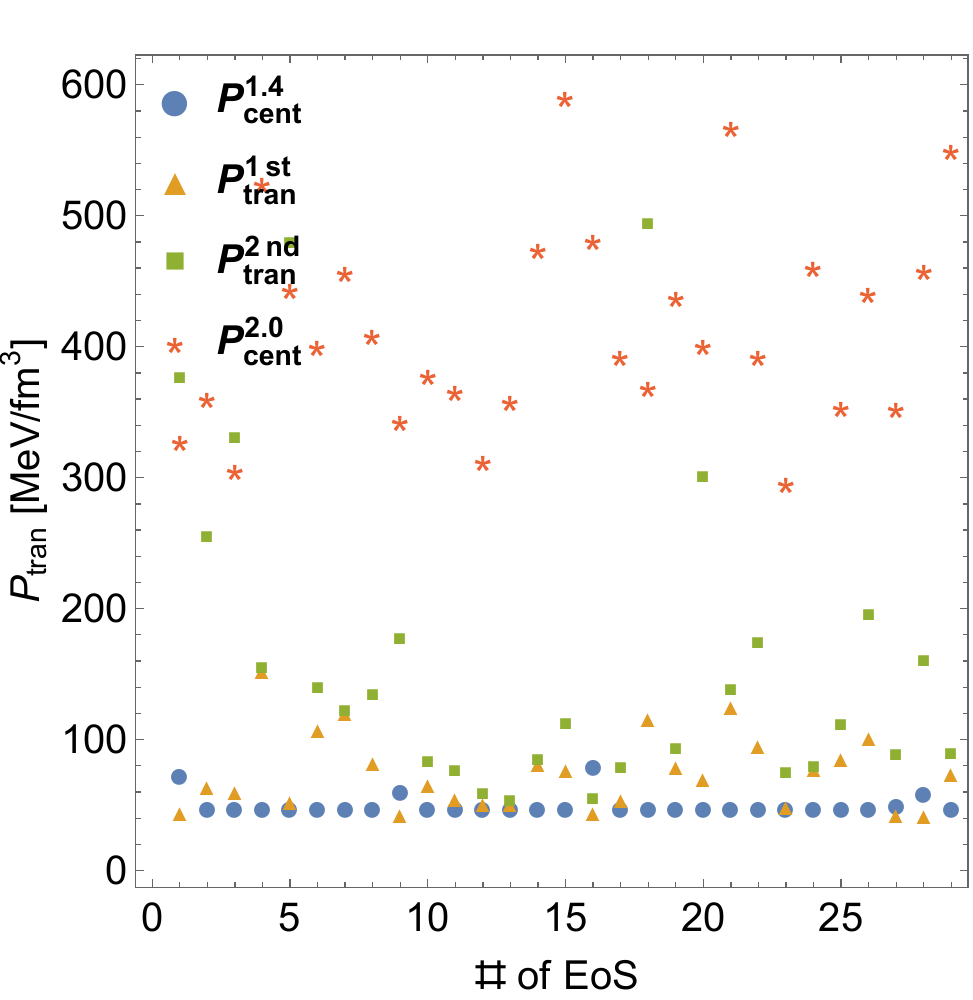} 
        \caption{We present the EoS, M-R curves, the $D_L$ and the transition pressure locations for 29 EoSs with first order phase transitions that have $D_L>1.5~km$. The red solid EoS and M-R curves in the top panels correspond to the five red points in the Fig. \ref{fig:Linear Correlation Distance}. The corresponding $D_L$ of these EoSs are presented in the lower left panel. In the lower right panel, the blue dot, orange triangle, green square and red star represent the central pressure of the 1.4 solar mass neutron star, the transition pressure of the 1st first-order phase transition, the transition pressure of the 2nd first-order phase transition, and the central pressure of the 2.0 solar mass neutron star of each EoS. }
        \label{fig:Linear Correlation example}
\end{figure*}

In Fig. \ref{fig:Linear Correlation example}, we presented 29 EoSs that have $D_L>1.5 ~\mathrm{km}$ out of 3,511 EoSs with 1st-order phase transitions, and their M-R curves and phase-transition features . The $D_\mathrm{L}$ of the first 5 EoSs are plotted as red points in Fig. \ref{fig:Linear Correlation Distance}. These EoSs are generated based on the DBHF EoS, but with random transition pressure and energy density discontinuity values attached to it. Consequently, they have exactly the same low-density EoS curve and low-mass M-R curve. However, due to the introduction of phase-transitions, their high-density EoS curve and M-R curve above 1.4 solar mass can be significantly different, as shown in the upper two panels of Fig. \ref{fig:Linear Correlation example}. In the bottom two panels of Fig. \ref{fig:Linear Correlation example}, we present the $D_\mathrm{L}$ and the transition pressure of these 29 EoSs. The EoSs have two first-order transitions. The second first-order phase-transition are likely less important in some cases, as we can see in the first and the fifth EoS (for example), the second transition pressure is above the central pressure of the 2.0 solar mass neutron star so that it cannot have any influence on the radius of a neutron star with $\leq$ 2.0 solar mass. The transition pressure of the first phase-transition of these EoSs, on the other hand, all interestingly locate around the central pressure of the 1.4 solar mass neutron stars. The phase transition happening here cannot have any significant influence on the radius of the 1.4 solar mass neutron star, but can easily influence the radius of the 2.0 solar mass neutron star, which results in a significant weakening of the radii correlation of two neutron stars. It suggests that the phase transitions sensitive to current NICER measurements may locate in between the central pressure of 1.4 and 2.0 solar mass neutron stars. In the future, with much more NS pairs to be analyzed, the $D_\mathrm{L}$ test could be used to test phase transitions over a broader range of NS densities. 

\bibliography{draft_mt.bib}{}
\bibliographystyle{aasjournal}

\end{document}